# ROTATING FRAME ANALYSIS OF RIGID BODY DYNAMICS IN SPACE PHASOR VARIABLES

**Shayak Bhattacharjee**

Department of Physics,
Indian Institute of Technology Kanpur,
NH-91, Kalyanpur,
Kanpur – 208016,
Uttar Pradesh, India.

\* \* \* \* \*

## CLASSIFICATION

## ABSTRACT

In this work we propose a new method of attacking problems in rigid body rotation, focussing in particular on the heavy symmetric top. The technique is a direct extension of the method traditionally applied to the free symmetric top. We write Euler's equations for the heavy top in a frame attached to it and thus sharing its entire angular velocity. The structure of the resulting equations is such that it is advantageous to cast and solve them in terms of complex variables (space phasors). Through this formalism, we obtain a direct link between the initial conditions at the time of launch and the subsequent behaviour of the top. The insertion of a damping term allows us to further explain the behaviour of a top where the pivot is non-ideal and has friction. Finally, we make some suggestions regarding experimental verification of our various results.

\* \* \* \* \*

# INTRODUCTION

A rather demanding part of any classical mechanics course is the section on rigid body rotation. The counter-intuitive behaviour of heavy tops and gyroscopes, together with the technical difficulties inherent in their analysis, has drawn the attention of teachers and researchers alike [1]-[4]. One very standard treatment of this problem is in Goldstein et. al. [5] where they construct the Lagrangian of the rotating top in terms of the Euler angles. Cyclic coordinates reveal first integrals of the motion and these can be used to reduce the problem to quadrature. This process does not make any assumptions or simplifications; however the quadrature cannot be integrated analytically and numerical methods must be used. Some conclusions can be drawn by observing the signs etc. of various quantities. Analytic representation of the dynamics, with quantitative expressions for the precession and nutation, can only be achieved in the limit of a fast top i.e. one where the spin kinetic energy far exceeds the gravitational potential energy. This approach is in fact similar to the one followed in Landau and Lifshitz [6] where the same fast top assumption is used for the analysis.

A second approach to the heavy symmetric top has been outlined in the text by Morin [7], which places greater emphasis than Goldstein on physical concepts and intuition. Here, a frame has been considered where the coordinate axes form a principal basis for the top. The frame does not share the top's entire angular velocity though but a part of it. The relation $\boldsymbol{\Gamma} = d\mathbf{L}/dt$ [8] is applied in this basis, and considerable algebraic manipulation is required to obtain the expressions for the time derivatives of the rotating axes. The result is a set of three coupled equations which describe the dynamics of spin, precession and nutation. As Morin has shown, the same equations can be obtained from the Lagrangian, with less effort. The solution of these equations, carried out under the assumption of a fast top, yields the precessional and nutational patterns. Yet another way of obtaining the same dynamical equations may be found in the engineering text by Merriam and Kraige [9] where Euler's equations have been employed in the principal basis used in Morin. This approach has also been adopted by Peraire et. al. [10] who analyse the consequences of the dynamic equations in much the same way as Morin.

A novel approach bypassing these equations has been proposed by Schoenhammer [11] where geometrical considerations are used to obtain the precession and nutation. The formal dynamical equations have been circumvented and instead an ingenious geometric construction has been used where components of angular momentum in various bases, including a non-orthogonal system, have been considered. The work has been extended and applied to more complicated problems such as the tippe top [12]. This treatment uses the reference frame mentioned above in connection with Morin's approach.

The standard procedure adopted by the above mentioned references for analysing the force free symmetric top is writing Euler's equations for it in its own frame, i.e. a frame which shares the entire angular velocity of the top. In this work we make a direct extension of this method to the heavy top, by writing Euler's equations for it in the body frame. It is here that we differ from the literature, where, as we mentioned above, the common practice is to consider a frame sharing only part of the top's angular velocity. Our choice of reference frame results in a structure which can be expressed elegantly in terms of certain complex combinations of the various angular velocities involved – these are the space phasors.

Sec. I of this paper presents the dynamical equations in their full generality. The following section deals with a few concrete applications of the proposed method. We first consider the free symmetric top and next, the heavy top. We then add on a damping term at the pivot and physically explain certain interesting behaviour displayed by the system. Finally we treat the case where the pivot is accelerating horizontally. We show that the answer obtained from our formalism agrees with physical intuition. We conclude the paper with a summary of the main results and some suggestions regarding experimental verification.

# I. EULER'S EQUATIONS IN SPACE PHASOR VARIABLES

The aim of this section is to write Euler's equations for a symmetric rigid body with no torque about its axis, and interpret the results in terms of the Euler angles with respect to a fixed lab frame.

We are familiar with Euler's equations of rigid body rotation i.e. $I_1(d\omega_1/dt)+(I_3-I_2)\omega_3\omega_2=\Gamma_1$ and its cyclic permutations but to avoid mistakes we give a brief note of how this equation is derived [13]. It is obtained by transformation from a body fixed frame (the principal basis) i.e. a rotating frame to an inertial frame which is oriented parallel to the body frame at the particular instant of time under consideration. This frame will hereafter be referred to as the body-aligned frame. For the transformation one has to use Coriolis' theorem,

$$\left(\frac{d\mathbf{L}}{dt}\right)_{\text{inertial}} = \left(\frac{d\mathbf{L}}{dt}\right)_{\text{rotating}} + \boldsymbol{\omega}\times\mathbf{L} \ , \tag{1}$$

and employ the fact that in a principal basis, $\mathbf{L}=(I_1\omega_1,I_2\omega_2,I_3\omega_3)$ [14].

Let us now list the potential choice of reference frames for the top. They are (*) the lab frame ($x,y,z$) which is fixed in space, the intermediate frame ($a,b,c$) and the body fixed frame ($d,q,o$) [15]. All the frames have a common origin – the top's pivot. The $x$-$y$-$z$ frame is fixed in space. The axes are orthogonal and gravity acts along the $-z$ direction as is conventional. Making an angle $\theta$ with the $z$ axis is the symmetry axis of the top which we denote by the $c$ axis. Considering the plane orthogonal to the $c$ axis, we let the $a$ axis be the line along which this plane intersects the horizontal $x$-$y$ plane. This $a$ axis is what is traditionally called the line of nodes. $\varphi$ is the angle in the $x$-$y$ plane between the $x$ and $a$ axes. Orthogonal to $a$ and $c$ is $b$ axis. Finally, the $d$ and $q$ axes are body fixed axes obtained from the $a$ and $b$ axes by a rotation in the $a$-$b$ plane through an angle $\psi$. This rotation takes place about the $c$ axis which remains invariant but for notational consistency we shall now refer to the same axis as the $o$ axis. Thus, $\theta$ becomes the angle of nutation, $\varphi$ that of precession and $\psi$ that of the top's spin. Descriptions of the axes are meaningless without a diagram and we show several views of the transformations in Figs. 1 and 2. We note that the rotations described here follow the $x$-convention of Goldstein, which is standard for rigid body problems.

We now write Euler's equations in the $d$-$q$-$o$ reference frame, using overhead dots for time derivative getting

$$I_d\dot\omega_d + (I_o-I_q)\omega_o\omega_q = \Gamma_d \ , \tag{2a}$$

$$I_q\dot\omega_q + (I_d-I_o)\omega_d\omega_o = \Gamma_q \ , \tag{2b}$$

$$I_o\dot\omega_o + (I_q-I_d)\omega_q\omega_d = \Gamma_o \ . \tag{2c}$$

To be absolutely correct, the above equation refers to the components of $\omega$ in the body-aligned frame and not the true body frame, hence they should be designated with subscripts different from $d$, $q$ and $o$ which refer to body fixed axes. Such a notation however would be pedantic, and we stick to the usual notation, while keeping this point in mind. If the top is symmetric $I_d$ will equal $I_q$ and if there is no torque about the $o$ axis, (2c) will at once lead to the conservation of $\omega_o$. We now see that under these conditions there is a symmetry of the remaining two equations which permits the use of the space phasor method.

The space phasor method is used to reduce the order of systems of real differential equations by introducing complex variables, called phasors. For instance, if there are two equations for two real variables, we will form a single phasor variable by adding $j$ (imaginary unit) times the second variable to the first. If the original system can be simplified using this ansatz, then the analysis is worthwhile. The technique finds application in physics in analysis of the Foucault pendulum [16] and in electrical engineering in the study of electric machinery [17,18].

(*) A note on the axis nomenclature : $x$-$y$-$z$ is of course pretty standard and needs no comment. We have intentionally not used symbols like $x'$-$y'$-$z'$ or $x_1$-$x_2$-$x_3$ for the other axes to highlight the fact that they belong to very different reference frames. The notation $a$-$b$-$c$ is clear; $d$-$q$-$o$ is taken from engineering terminology where $d$ represents "direct" and $q$ "quadrature" meaning "at right angles". $o$ is for "zero-sequence"; a technicality perhaps confusing to the reader who can very reasonably imagine the letter to stand for "orthogonal".

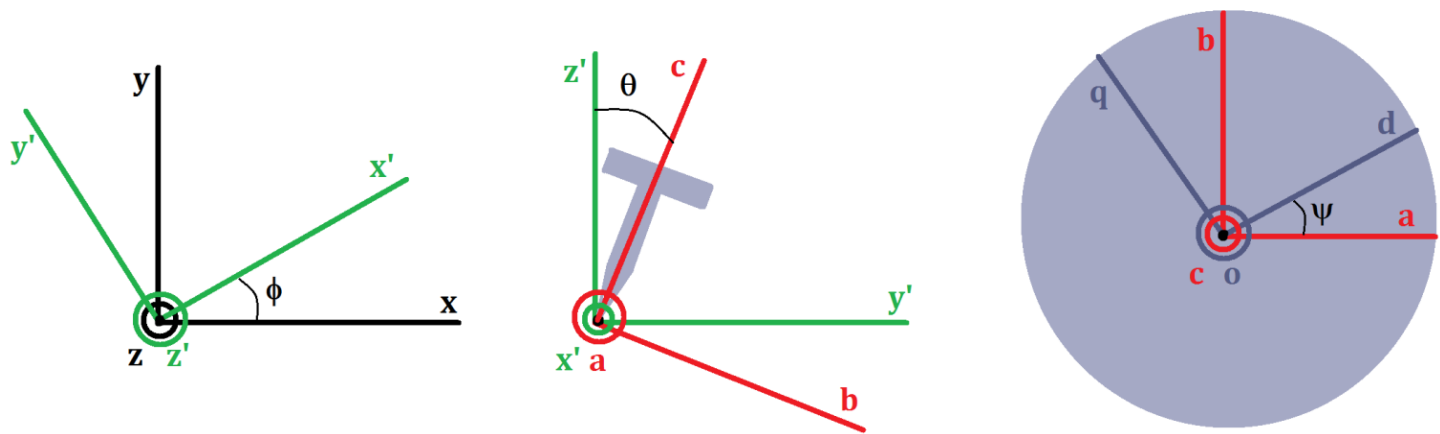

Figure 1 : *Orthographic views of orthogonal transformations. In each view the axis of rotation comes out of the plane of the paper and is shown as a circle in the relevant colour. The x-y-z system is fixed in the lab frame with gravity along the –z direction. The origin is at the top's pivot. The first rotation (left panel) is through angle φ about the z axis to form the x'-y'-z' axes (these axes are not used in the analysis that follows). The next step (middle panel) consists of a rotation through angle θ about the x' axis and the set of axes thus formed is denoted by a-b-c. The axis of symmetry of the top lies along the c axis and the top can be seen in profile view. The final step (right panel) is a rotation through angle ψ about the c axis to get the d-q-o axes. The coloured background highlights the fact that the rotation takes place in the plane of the top's spin. The slate grey colour has been consistently used to indicate the top in this and the next figure; moreover the d q and o axes are shown in a darker shade of the same colour to highlight the fact that they are fixed in the top's frame. We note that a-b-c and d-q-o are both principal axes for the top – however the latter frame carries the entire angular velocity of the top while the former has only a part of it.*

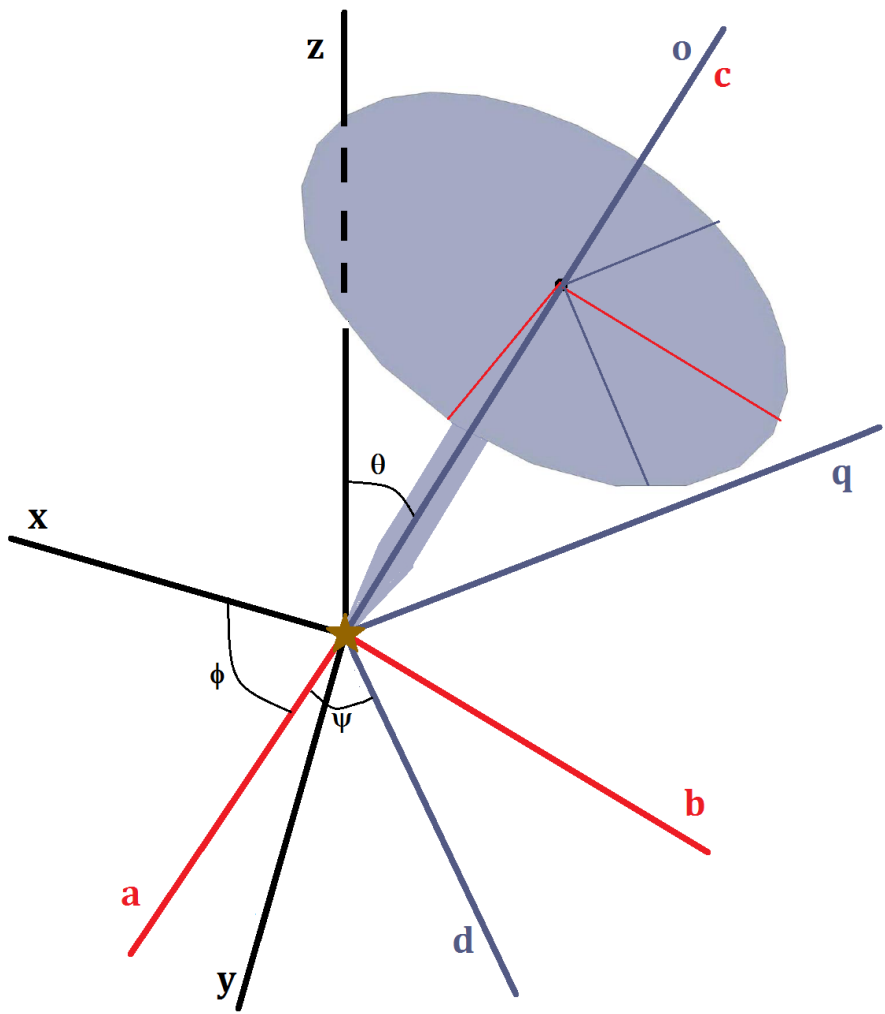

Figure 2 : *Three-dimensional view of the top. The axes and angles mentioned in Fig. 1 can be seen clearly. We should remember that the rotations through φ and ψ are not coplanar. As written in the text, the rotation is per the x-convention of Goldstein.*

Letting $I_d = I_q = I$ and using the conservation of $\omega_o$, we multiply (2b) by $j$ and add it to (2a) getting

$$\frac{\mathrm{d}}{\mathrm{d}t}\left(\omega_d + j\omega_q\right) + j\frac{I - I_o}{I}\omega_o\left(\omega_d + j\omega_q\right) = \frac{1}{I}\left(\Gamma_d + j\Gamma_q\right) \quad . \tag{3}$$

Quite clearly, a space phasor formulation is possible in terms of the phasor $\omega_d + j\omega_q$ which we denote by $\underline{\omega}$.

We now make the definition

$$v = \frac{I - I_o}{I}\omega_o \quad . \tag{4}$$

For the case considered in the rest of this paper i.e. a symmetric rigid body with no torque about the $o$ axis $v$ is a constant of the motion. Letting $\underline{\Gamma} = \Gamma_d + j\Gamma_q$ we obtain the space phasor form of (3) as

$$\frac{\mathrm{d}}{\mathrm{d}t}\underline{\omega} + jv\underline{\omega} = \frac{1}{I}\underline{\Gamma} \quad . \tag{5}$$

This is the equation which must be solved for the dynamics of the body frame angular velocities. Its structure arises from our choice of applying Euler's equations in the *d-q-o* frame. The space phasors are primarily a calculational convenience and (5) could equivalently have been written as two equations, one for the real and one for the imaginary part. However, it turns out that the manipulation is easier in terms of the phasor variables, which is why we use them throughout this paper.

Equation (5) thus gives the dynamics of the top as it would appear in the body-aligned frame. Until this point, our approach mirrors the standard formulation of the problem of the free symmetric top, which is then analysed in the body-aligned frame itself. However, for the heavy symmetric top, a formulation in terms of the easily visualizable precession, nutation and spin is desired. For this reason, the treatments [7,9,10] which do employ Euler's equations for the heavy top go over to the *a-b-c* frame. However, we now plow straight on and attempt to derive the relation between the components of angular velocity in the body-aligned frame and those in the lab frame. Since both the body-aligned frame and the lab frame are inertial, we can switch from one to the other using a simple geometric transformation through the Euler angles. These angles of course will vary in time as the top rotates. We now need the transformation relations between the various sets of axes. Before presenting the rotation matrices we mention that a formal knowledge of the theory of rotations is absolutely not required for understanding this paper. The transformations have been shown very clearly in Fig. 1 and it is an easy exercise in trigonometry to derive the matrix from the diagram. The transformation from the *a-b-c* to the *d-q-o* basis is simple:

$$\begin{bmatrix}\hat{\mathbf{d}}\\\hat{\mathbf{q}}\\\hat{\mathbf{o}}\end{bmatrix} = \begin{bmatrix}\cos\psi & \sin\psi & 0\\-\sin\psi & \cos\psi & 0\\0 & 0 & 1\end{bmatrix}\begin{bmatrix}\hat{\mathbf{a}}\\\hat{\mathbf{b}}\\\hat{\mathbf{c}}\end{bmatrix} \quad . \tag{6}$$

We note that $\theta$, $\varphi$ and $\psi$ are functions of $t$. We quote from Goldstein the formula for the complete *x-y-z* to *d-q-o* transformation :

$$\begin{bmatrix}\hat{\mathbf{d}}\\\hat{\mathbf{q}}\\\hat{\mathbf{o}}\end{bmatrix} = \begin{bmatrix}\cos\varphi\cos\psi - \cos\theta\sin\varphi\sin\psi & \sin\varphi\cos\psi + \cos\theta\cos\varphi\sin\psi & \sin\theta\sin\psi\\-\cos\varphi\sin\psi - \cos\theta\sin\varphi\cos\psi & -\sin\varphi\sin\psi + \cos\theta\cos\varphi\cos\psi & \sin\theta\cos\psi\\\sin\theta\sin\varphi & -\sin\theta\cos\varphi & \cos\theta\end{bmatrix}\begin{bmatrix}\hat{\mathbf{x}}\\\hat{\mathbf{y}}\\\hat{\mathbf{z}}\end{bmatrix} \quad . \tag{7}$$

Geometrical considerations yield the directions of the various angular velocities : $\dot{\theta}$ is along the *a* axis, $\dot{\varphi}$ along the *z* axis and $\dot{\psi}$ along the *o* axis. These are obvious from the orthographic views of Fig. 1. Now projecting these onto the *d*, *q* and *o* axes yields the corresponding components of the angular velocity as follows.

$$\omega_d = \dot{\theta}\cos\psi + \dot{\varphi}\sin\theta\sin\psi \quad , \tag{8a}$$

$$\omega_q = -\dot{\theta}\sin\psi + \dot{\varphi}\sin\theta\cos\psi \quad , \tag{8b}$$

$$\omega_o = \dot{\psi} + \dot{\varphi}\cos\theta = \text{const.} \quad . \tag{8c}$$

Since the main equation (5) deals with the space phasor we combine (8a)+*j*(8b) to get

$$\underline{\omega} = \left(\dot{\theta} + j\dot{\varphi}\sin\theta\right)\underline{1}e^{-j\psi} \quad . \tag{9}$$

The **1** on the right hand side (RHS) of the above refers to the unit phasor 1+*j*0. It is a technical construct and has been introduced to maintain consistency so that both sides of the equation become phasors. Otherwise the entire coefficient of the exponential would have to be marked as a phasor, which is cumbersome notation. In practice one gets by pretty well without writing the **1** (I did not while doing these calculations) and in an examination, the candidates should not be penalised for omitting it.

It can be checked that substituting (9) into (5) and equating real and imaginary components, one recovers the usual dynamical equations for the top i.e. Eq. 9.70 of Morin. The advantage of our derivation is that it is physically illuminating and requires a minimum amount of algebraic manipulation. Also, there is a distinct gain in the ease of finding solutions as one of our equations i.e. (5) is linear and only the second one is nonlinear. We note that solving (5) and (9) together give us the complete behaviour of a symmetric rigid body in terms of nutation, precession and spin. What we have done until this point is simply a mathematical restatement of the formulation of the rigid body problem – we have not obtained a solution ! Also, we have not made any assumptions, approximations or simplifications in this derivation which is thus fully general. The approximations will come once we start using the method to obtain analytically tractable solutions.

## II. ILLUSTRATIVE EXAMPLES

Before we begin this section we make two assumptions which facilitate the calculations. First, we assume that the top always stays close to the vertical so that sin$\theta$≈$\theta$ and cos$\theta$≈1. Next, we assume that the top is fast i.e. the spin speed about the *o* axis is very high. Very high means that as the top precesses and nutates, there will be very little change in the spin speed. Because of this, the spin rate can be treated as a constant which we call Ω. Hence we have the relation $\psi = \Omega t$. This of course is the fast top assumption which we have already mentioned in the Introduction. The approximations may appear somewhat restrictive, but the incredible technical difficulty of the problem under consideration makes such assumptions necessary for obtaining actual analytical solutions. Morin mentions these simplifications quite explicitly before obtaining the general nutating solutions (Eq. 9.84 onwards). Goldstein initially gives a semi-qualitative discussion based on the reduction to quadrature where these approximations are not made. However, he does take recourse to these simplifications while presenting quantitative expressions for the frequency of precession and nutation and the amplitude of the nutation (Eq. 5.65 onwards). Landau too makes the fast top assumption for deriving the precession frequency. Since our aim here also is to make quantitative predictions, we too must assume that the top is fast. To some extent, the assumptions can be justified by looking at the main practical application of rotational dynamics – the gyroscope. In this application it is essential that the spin speed be very high so the fast top assumption automatically holds. As mentioned in all the texts, and as we will also see for ourselves, the nutation amplitude decreases as the spin speed increases, and hence the small angle approximation can be said to be a consequence of the fast top assumption.

We note that all the calculations of this section will use these approximations throughout so in future we will not bother to mention them explicitly every time we use them. More accurate results may be obtained through a perturbative treatment, however such a discussion in this paper will take us too far afield.

## 1. FREE SYMMETRIC TOP (FST)

This is the easiest case – the torque on the top is simply zero and we have from Euler's equations in the form of (5),

$$\frac{d}{dt}\underline{\omega} + jv\underline{\omega} = 0 \quad . \tag{10}$$

At this point we could have proceeded along the usual lines and simply made a body-aligned frame analysis without bothering to find the behaviour in the lab frame. This analysis would have been exact, requiring no assumptions or simplifications as (10) is a linear equation. However, we will see later that the FST solutions are in fact the homogeneous solutions of the heavy top, where we certainly do want the lab frame dynamics. Hence we use (9) to connect between the body-aligned inertial frame and the lab frame. Since (9) is heavily nonlinear we now need to make the fast top and the small nutation approximations to linearize it and make the analysis tractable. This is just what is seen in the literature – the assumptions and approximations come in only when one wants results in terms of the Euler angles. Solving (10) and using (9) together with the above mentioned approximations we readily get

$$\left(\dot{\theta} + j\theta\dot{\varphi}\right)\underline{1} = \underline{\omega}_{00} e^{j(\Omega-v)t} \quad , \tag{11}$$

where $\underline{\omega}_{00}$ denotes $\omega$ at $t=0$ (the double zero is to avoid confusion with $\omega_o$). We note that $\Omega$, though (approximately) constant in time, is not a constant of the motion, like $v$. Initial conditions determine the value of $v$ while $\Omega$ will also depend on the speed of precession. Hence the quantity $\Omega-v$ does not in fact have a definite *a priori* value and becomes determinate only when set equal to a known quantity. Now the real and imaginary parts of (11) must separately be equal, and the left hand side (LHS) is easy to split into these parts as $\theta$ and $\varphi$ are both real. Let us examine the class of solutions featuring precession without nutation. This requires the LHS of (11) to be purely imaginary throughout, and at once forces the exponent on the RHS to be identically zero. If that were not so, the real part of the RHS would fluctuate in time, contradicting the condition of no nutation. Since $v = (I-I_o)\omega_o / I$ and $\omega_o = \Omega + \dot{\varphi}$, we have the precession frequency from

$$\Omega - v = 0 \Rightarrow \dot{\varphi} = \frac{I_o \Omega}{I - I_o} \quad . \tag{12}$$

This result agrees with the standard FST results found in books, which one can verify using the formula for $\omega_o$. This condition alone is clearly not sufficient to guarantee regular precession as we must also have $\underline{\omega}_{00}$ purely imaginary. This is reasonable as it means that an initial impulse in the $\varphi$ direction is necessary to initiate nutation-free precession.

We now analyse a more general case. Suppose the initial condition features impulses in both $\theta$ and $\varphi$ directions such that $\underline{\omega}_{00} = \omega_{01} + j\omega_{0j}$. Then applying (11) and separating real and imaginary parts yields

$$\dot{\theta} = \omega_{01}\cos(\Omega-v)t - \omega_{0j}\sin(\Omega-v)t \quad , \tag{13a}$$
$$\theta\dot{\varphi} = \omega_{01}\sin(\Omega-v)t + \omega_{0j}\cos(\Omega-v)t \quad . \tag{13b}$$

Right now no guesses can be made as regards the functional nature of $\theta$ and $\varphi$ so we adopt a systematic approach. We see that (13a) features only one variable while (13b) features two. Hence we first integrate (13a) to yield

$$\theta = C + \frac{\omega_{01}}{\Omega-v}\sin(\Omega-v)t + \frac{\omega_{0j}}{\Omega-v}\cos(\Omega-v)t \quad . \tag{14}$$

Now (13b) will yield $\dot{\varphi}$ on division by (14). For arbitrary $C$, the quotient does not have a clean form. However, the special value $C=0$ will produce a neat expression for $\dot{\varphi}$ so a detailed investigation of $C$ is warranted. Its value is clearly dependent on the initial conditions through the simple relation $\theta_{00} = C + \frac{\omega_{0j}}{\Omega - \nu}$, where $\theta_{00}$ is of course the launch angle. The key realization is that because the top is free, the lab frame can be chosen arbitrarily. The dynamic equations are obtained from consideration of vectors which are independent of the coordinate system chosen, and the Euler angles can be defined so long as the lab set consists of an orthonormal basis. Hence none of the preceding calculations refer in any way to any particular orientation of the lab set, which we call *L*1-*L*2-*L*3. The lab axes can be oriented any which way in space without affecting $\omega_{01}$ and $\omega_{0j}$, which are determined uniquely the moment the body frame is fixed. However, different choices of the lab frame axes will lead to different values of $\theta$ and $\varphi$ (and $\psi$) corresponding to any particular orientation of the top. It is thus possible to choose the lab frame such that the value of $\theta$ corresponding to the top's initial position equals $\frac{\omega_{0j}}{\Omega - \nu}$. For example, consider the top of Fig. 2 placed in a force free situation in the orientation shown in that figure. Because of the free nature of the motion the particular lab axes *x-y-z* have no individual significance. Let the *d-q-o* axes be as shown in the figure, and suppose that at launch the top is given an angular velocity only in the **d** direction (other than spin of course). Then $\omega_{0j}$ will be zero, and the dynamics will have a tractable form in a lab frame where $\theta_{00}$ is zero. In other words, the correct lab frame will have its *L*3 axis oriented with the initial configuration of the *o* axis. In this specially chosen lab frame, (14) will factor out of (13b) leaving uniform precession. At once the quantity Ω-ν becomes determinate and setting it equal to $\dot{\varphi}$ we have the precession frequency

$$\dot{\varphi} = \frac{I_o \Omega}{2I - I_o} \quad . \tag{15}$$

The equation for $\theta$ now gives the nutation

$$\theta = \frac{2I - I_o}{I_o \Omega} \left( \omega_{01} \sin \frac{I_o \Omega t}{2I - I_o} + \omega_{0j} \cos \frac{I_o \Omega t}{2I - I_o} \right) \quad . \tag{16}$$

The trajectory of the tip of the top described by (15) and (16) in terms of $\theta$ and $\varphi$ defined with respect to *L*1-*L*2-*L*3 is shown in Fig. 3. For comparison, a set of primed axes is also indicated, showing the difficulties involved in characterizing the motion with respect to a wrongly aligned set of axes. From (16) we recover the result expressed in Goldstein and Morin that the nutation frequency varies directly as the top's spin rate, and the nutation amplitude shows an inverse relationship with it.

We also see that the nutation frequency tends to blow up for a particular ratio between the two moments of inertia namely $I_o$:$I$=2:1 – we note that this is the largest possible value the ratio can have. To increase the ratio, we must have as little mass distribution along the *o* axis as possible and circular symmetry forces the shape of the top to be a solid of revolution. The two criteria indicate that the shape of top which maximises the ratio is the plane disk – then too the ratio of 2 is achieved if the pivot be the same as the centre of mass – not a realistic top at all. Any deviation from this ideal shape only tilts the balance in favour of *I* (which had better be so).

Our method has thus yielded the FST solutions in the lab frame and we move on to the heavy symmetric top which we have already mentioned is the focal point of our paper.

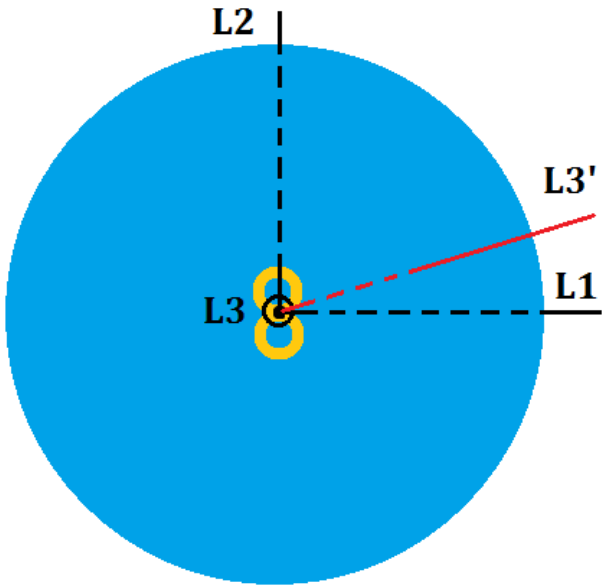

Figure 3 : *This figure shows the trajectory of the tip of a top satisfying (15) and (16) with the specific set of initial conditions mentioned in the text. Obviously, the trajectory lies on the surface of a sphere. Here, the sphere in shown in blue and the trajectory in yellow. The lab frame axes* L1, L2 *and* L3 *are the suitable axes in which* C=0 *and the motion has a simple description. They have their origin at the top's pivot.* L1 *and* L2 *lie in the equatorial plane, hence seen as dashed lines.* L3 *comes directly out of the plane through the line of viewing, because of which it appears in point view.* θ *and* φ *are defined with respect to* L1-L2-L3. *A second choice of lab frame axes has been denoted by primes. Of these,* L3' *is shown as an axis which makes a certain inclination with* L3. *The trajectory is very difficult to describe in terms of Euler angles defined in the primed set.*

## 2. HEAVY SYMMETRIC TOP (HST)

We now take on the HST using our formalism. Fig. 1 (centre panel) shows that the torque of gravity acts along the *a* axis; in fact along the $-\hat{\mathbf{a}}$ direction as it is the cross product of the displacement ($\hat{\mathbf{c}}$) and the force component($\hat{\mathbf{b}}$). The magnitude of the torque is $Mgh\theta$ where *h* is the distance along the *o* axis from the pivot to the centre of mass of the top. The **Γ** phasor is obtained by applying (6) on this torque, and substituting this into (5) yields

$$\frac{\mathrm{d}}{\mathrm{d}t}\underline{\omega} + j\nu\underline{\omega} = -\frac{Mgh\theta}{I}\underline{\mathbf{1}}e^{-j\Omega t} \quad . \tag{17}$$

This equation of course has homogenous and particular solutions – the homogeneous ones are nothing but the FST solutions, (10)ff. and the particular solutions are the ones on which we will focus now. This solution is, using subscript *p* for particular, and assuming $\theta_p$ to vary slowly in comparison with $\Omega t$:

$$\underline{\omega}_p = j\frac{Mgh\theta_p}{I(\Omega-\nu)}\underline{\mathbf{1}}e^{-j\Omega t} \quad , \tag{18}$$

on which (9) is applied to yield

$$\dot{\theta}_p + j\theta_p\dot{\varphi}_p = j\frac{Mgh\theta_p}{I(\Omega-\nu)} \quad . \tag{19}$$

The RHS of the above is purely imaginary so we at once have $\dot{\theta}_p = 0$ whereby $\theta_p$ becomes a constant. Hence our slowly varying *θ* assumption leads to no error after all. Equating the imaginary parts yields the precession frequency; the equation is implicit however as *v* itself has a contribution from $\dot{\varphi}$. Writing things explicitly we recover a quadratic for $\dot{\varphi}$:

$$(I_o - I)\dot{\varphi}^2 + I_o\Omega\dot{\varphi} + Mgh = 0 \quad , \tag{20}$$

which has the usual slow and fast precession frequency solutions.

We now show that our formalism provides a direct link between the release conditions and the top's subsequent behaviour. This in fact is a topic which is often not discussed explicitly in books on mechanics – they give the dynamical equations but then do not mention what the top will actually do if say released from rest at an angle or released from vertical with an angular impulse. This aspect of the problem is ideally suited to our approach.

There are only three release parameters of a HST which make physical sense. They are the launching angle $\theta$, and the launch time angular velocities in the $\theta$ and $\varphi$ direction. $\varphi$ is a cyclic or ignorable coordinate for the HST and we can choose our axes such that its initial value is zero. Let us say the initial launching conditions are $\theta_{00}$, $\dot{\theta}_{00}$ and $\dot{\varphi}_{00}$. As before we have used the double nought subscript to avoid confusion with $o$ axis components. Now $\left[\dot{\theta}_p\right]_{t=0} = 0$ and $\left[\dot{\varphi}_p\right]_{t=0} = \dot{\varphi}_{\text{fast}}$ or $\dot{\varphi}_{\text{slow}}$ depending on the precession mode which the top chooses to go into. Generally it is the slow mode which is preferred, although it is not relevant here. The difference between the initial values and the $t=0$ particular solutions gives the $t=0$ homogeneous solutions (subscript $h$). But we know that these evolve according to (11):

$$\left(\dot{\theta}_h + j\theta_h\dot{\varphi}_h\right)\mathbf{1} = \left(\left[\dot{\theta}_h\right]_{t=0} + j\theta_{00}\left[\dot{\varphi}_h\right]_{t=0}\right)\mathbf{1}\,e^{j(\Omega-\nu)t} \quad . \tag{21}$$

And in the previous section we showed that (21) can always be solved, if need be by reorienting the lab frame axes. Of course, this time the *x-y-z* axes are not arbitrary as gravity makes the space anisotropic. However, axis reorientation is eminently permissible while treating the homogeneous torque-free solutions. Thus the particular solutions are always obtained in the fixed lab frame with *z* along gravity, while the homogeneous solutions may be derived in an inertial frame making an inclination with it. Some axis transformations will then be required to change from the reoriented axes to the original ones and thus obtain the precession and nutation in the gravity-aligned basis. As a qualitative example once again consider the scenario treated as a demonstration of how to apply (13) – (16). This time let the top in Fig. 2 be a heavy one with gravity along the *–z* axis. Suppose the initial conditions are such that the proper basis for the homogenous motion is the *L*1-*L*2-*L*3 basis of Fig. 3, with the *L*3 axis aligned along the initial *o* axis. Then the homogeneous motion will be described by the loops shown in Fig. 3, while the particular solution will cause the top to precess around the *z* axis. Since the amplitude of the nutational motion is very small, the effect of the particular integral will dominate and the homogeneous solution will manifest itself as nutation and also as a periodic fluctuation of the precessional frequency. This periodic perturbation is what Goldstein has mentioned in his discussion of pesudoregular precession – the average precessional frequency is unchanged but at a microscopic level there are variations, periodic with the nutation frequency.

One of the demerits of the standard formulation of the top problem is that the effects of friction at the pivot are neglected. As Goldstein comments, "the effects of friction....cannot be directly included in the Lagrangian framework". Nevertheless he also says that, for a practical top, "the nutation is damped out by friction at the pivot..." and "the top then appears to precess uniformly". This statement is in fact not intuitive as one may ask why the friction selectively damps out the nutation leaving the precession unharmed. Damped motion of a FST has been considered in [19] based on geometrical arguments. We see that our approach is amenable to the insertion of a friction term and the analysis of this system is the focus of the next subsection.

## 3. DAMPED COMPENSATED (DC) TOP

In many practical tops the pivot is non-ideal in that there is friction between it and the ground. Such a frictional force will have a complex form dependent on the instantaneous normal reaction, the nature of motion (sliding, rolling or boring) etc. This is very difficult to model mathematically. As a first approximation however, one can replace it with a damping torque, proportional to the angular velocity. The body frame angular velocity components are the natural choice as that is the only significant frame from the viewpoint of the top (the lab frame becomes special only if gravity is present but the simple friction of our model acts on the FST too). Hence we make the insertions of the form

$$\Gamma_d = -\gamma_1 \omega_d \quad , \tag{22a}$$

$$\Gamma_q = -\gamma_2 \omega_q \quad , \tag{22b}$$

$$\Gamma_o = -\gamma_3 \omega_o \quad , \tag{22c}$$

where the $\gamma$'s (all greater than zero) denote damping coefficients which need not be equal on the three axes. But common sense says that they will be equal for the $d$ and $q$ axes and much less for the $o$ axis. For instance, a crude gyroscope may use a ball bearing for mounting the rotor on its own axis but may use an ordinary (sliding) pivot to make contact with the ground. More sophisticated gyros will use a rotor powered by a motor which will produce a torque to counter the friction and there will be no effective damping on the $o$ axis. Damping on the $o$ axis is anyway an undesirable phenomenon as it will be a hindrance in experimentation using the gyro. Hence we assume that $\gamma_1 = \gamma_2 = \gamma$ and $\gamma_3 = 0$. This last property is what we refer to by "compensation" in the title of this subsection – the $o$ axis damping is compensated by an external motor. Accordingly we have the modified form of (5) :

$$\frac{\mathrm{d}}{\mathrm{d}t}\underline{\omega} + (\gamma + j\nu)\underline{\omega} = \frac{1}{I}\underline{\Gamma} \quad . \tag{23}$$

This is the governing equation of the DCT. The homogeneous solution this time decays quickly to zero. The damped compensated FST (DCFST) will simply come to rest at some position spinning away without precessing or nutating, and this does not make for a worthwhile study. The damped compensated HST (DCHST) is more interesting and we start by writing its characteristic equation

$$\frac{\mathrm{d}}{\mathrm{d}t}\underline{\omega} + (\gamma + j\nu)\underline{\omega} = -\frac{Mgh\theta}{I}\underline{1}\mathrm{e}^{-j\Omega t} \quad . \tag{24}$$

Since the homogenous solutions are damped out we focus only on the particular integral, this time omitting the subscript $p$ which we had used in Subsection II.2. Solving (24), applying (9) and separating real and imaginary gives the following equations :

$$\dot{\theta} = \frac{-\gamma Mgh}{I(\gamma^2 + (\Omega-\nu)^2)}\theta \quad , \tag{25a}$$

$$\dot{\varphi} = \frac{(\Omega-\nu)Mgh}{I(\gamma^2 + (\Omega-\nu)^2)} \quad . \tag{25b}$$

Like (19) the equation for the precession is implicit and must be unscrambled to form a simple cubic; however that is not really too relevant. The presence of a damping term suggests that a fast precession state will not be favoured. For slow precession we note that the term $\Omega-\nu$ evaluates to $I_o\Omega/I$. Now (25a) yields that the polar angle starts out from its initial value and decays to zero with a certain time constant. Making the assumption of slow precession, this time constant evaluates to

$$\tau = \frac{I^2\gamma^2 + I_o^2\Omega^2}{\gamma IMgh} \quad . \tag{26}$$

We note that this is quite different from the time constant $1/\gamma$ with which the homogeneous solutions die out. In fact for a very fast top, the homogeneous solutions are killed off much quicker than the time it takes for the angle to vanish. As the nutation arises only from the homogeneous solution [(25a) does not describe nutation under any circumstances] we have mathematically explained Goldstein's comment that the nutation dies out and the top appears to precess uniformly.

Another interesting feature of the solution is that the polar angle slowly but surely goes to zero. For very small $\theta$, the precession is meaningless as it is scarcely visible. Hence the eventual state of the top is to spin uniformly pointing steadily in the upward vertical direction. It is a classic display of levitation – the top simply stands in equilibrium at the position which gravity would have considered the most unstable, mocking gravity to bring it down if it can. This phenomenon is similar to what happens in a tippie top [12]. The ease of obtaining this solution encourages us to apply the method to a still more complicated case where the pivot is given an acceleration in the horizontal plane. That is the topic of the next subsection.

## 4. DCHST WITH HORIZONTAL ACCELERATION

In this subsection we consider the DCHST where the pivot experiences a uniform horizontal acceleration ($a$ in the $-\hat{\mathbf{x}}$ direction to be specific, not to be confused with the similar label of the axis, which does not appear in this subsection). As in the previous section we are interested in the long time state. Given the previous results, this problem is in fact a one-liner : the final state is no precession and no nutation, $\theta = \tan^{-1}(a/g) \approx a/g$ and $\varphi = \pi/2$. In the pivot frame the centre of mass (CM) of the top experiences a pseudo force $+Ma$ in the $\hat{\mathbf{x}}$ direction which is entirely equivalent to the gravitational force. Now the choice of the lab frame axes is arbitrary and we will recover the previous problem *in toto* if we align $-z$ with the direction of the net force on the CM. The final state of the top would be to point along this new $+z$ which in this case means vertically upward and in the direction of the acceleration. An easy check shows that the new $z$ axis does indeed make the angles we claimed above with the original axes, and the problem is solved.

In this paper however we are trying to see the applicability of our method so we prove that we can derive this result in our original *x-y-z* basis. The pseudo force on the CM is $Ma\hat{\mathbf{x}}$ and the torque is obtained by $\mathbf{r} \times \mathbf{F}$ as usual. Now for the *d-q-o* components of torque we must expand $\mathbf{F}$ in the *d-q-o* basis and for this purpose we use the first column of the transformation matrix (7). Evaluating the cross product we have

$$\mathbf{\Gamma} = Mah\left[\left(\sin\varphi\cos\psi + \cos\varphi\sin\psi\right)\hat{\mathbf{d}} + \left(\cos\varphi\cos\psi - \sin\varphi\sin\psi\right)\hat{\mathbf{q}}\right] \quad . \tag{27}$$

Recognizing the sines and cosines of sums, and adding on the gravitational torque we construct the $\underline{\mathbf{\Gamma}}$ phasor and the governing equation for the top is

$$\frac{d}{dt}\underline{\omega} + \left(\gamma + j\nu\right)\underline{\omega} = \frac{Mh}{I}\left(g\theta + jae^{-j\varphi}\right)\underline{\mathbf{1}}e^{-j\Omega t} \quad . \tag{28}$$

The problematic term here is the exp(-$j\varphi$) on the RHS. But we can save a lot of messy algebra by recognizing that the slow precessional states are favoured for the damped top. Under this condition the time variation of the term involving $\varphi$ will be much less than that of the term featuring $\Omega t$. This is the same argument as the one we used to treat $\theta$ as a constant in (18); as it happened we accrued no error in that case. Here, the error will not fortuitously evaluate to zero but we can definitely integrate (28) at least for a few cycles of spin, assuming $\varphi$ to be a constant. This simplified integral may not be an exact description of the motion but let us see how much information we can get out of it. The particular solution with this assumption is

$$\underline{\omega}_p = \frac{-Mgh\theta + jMahe^{-j\varphi}}{I\left[\gamma - j\left(I_o\Omega/I\right)\right]}\underline{\mathbf{1}}e^{-j\Omega t} \quad , \tag{29}$$

where we have written the form of $\Omega$-$\nu$ appropriate for slow precession.

Using (9) and separating the real and imaginary parts yields

$$\dot{\theta} = \frac{1}{I\left(\gamma^2 + (I_o/I)^2\Omega^2\right)}\left[-\gamma Mgh\theta + \gamma Mah\sin\varphi - \frac{I_o\Omega}{I}Mah\cos\varphi\right] \quad , \tag{30a}$$

$$\theta\dot{\varphi} = \frac{1}{I\left(\gamma^2 + (I_o/I)^2\Omega^2\right)}\left[-\frac{I_o\Omega}{I}Mgh\theta + \frac{I_o\Omega}{I}Mah\sin\varphi + \gamma Mah\cos\varphi\right] \quad . \tag{30b}$$

These equations are difficult to solve in detail and the easiest option is to perform at this stage the axis rotation which we could have executed right at the beginning. The fixed point of the dynamics is evident though on inspection :

$$\varphi = \pi/2; \theta = a/g; \dot{\theta} = \dot{\varphi} = 0 \quad , \tag{31}$$

which we have already shown is the true fixed point of the system.

This at last completes the set of examples which we had wanted to demonstrate using our formalism. We close with a brief section where we summarise the main ideas and give some details regarding experimental visualization of the various conclusions we have drawn.

# RECAPITULATION AND CONCLUSION

The main point of this paper has been the proposal of a new technique of handling problems on rigid body rotation. We have selected the body frame as the reference frame of choice and used Euler's equations (2) in it. The assumptions of symmetry and of no torque about the figure axis have resulted in the space phasor method being applicable on (2) to produce (5) which is one of our central equations. The conversion from the abstract body frame angular velocities to the concrete, observable ones has been done using axis transformations (7), resulting in the second key equation (9). Until this point (i.e. in Sec. I) we have made no simplifications but merely done a mathematical reformulation of the top problem. The usual system of two second order differential equations with somewhat messy non-linear terms has been converted to a system of two nested complex equations of which the first one is linear. We say "nested" as (5) is like the outer equation describing the evolution of the body-aligned frame angular velocities and (9) the inner equation describing the connection between the body-aligned angular velocities and the observable motions.

The first application of the formalism is to the free symmetric top where the body frame analysis mirrors the standard treatment. The conversion to lab frame angular velocities has also been shown. The importance of this section is primarily because the free top solutions are in fact the homogeneous solutions of the heavy symmetric top. This homogeneous-particular approach is a direct method of determining the relation between the launch conditions of the heavy top and its subsequent motion. The particular solutions of (9) for the heavy top are given in (19); they admit uniform precession at a choice of two different frequencies, and no nutation. The $t=0$ homogeneous solutions are obtained from the initial conditions and their evolution is governed by (21). This may have to be solved in an inertial frame making a finite inclination with the gravity-aligned lab frame. The solutions allow for nutation and pseudoregular precession.

Verification of the predictions of (19) and (21) needs a system where frictional effect is very low. One such experimental setup is a variation of the apparatus mentioned in Kleppner [20]. Our pivot consists of a large, light spherical ball which is kept in a closely fitting cup. Now the cup is connected to an air blower so that the ball does not actually make contact with the cup but rides on an air cushion a few millimetres above it. This top can be subjected to various initial conditions and the agreement with our predictions can be measured.

For the damped compensated top we see that the homogeneous solutions which include fast precession and nutation decay exponentially in time. The particular solution features uniform precession at a choice of three frequencies but because of the damping term, the slow precession is expected to be prevalent. The polar angle also decreases exponentially in time and eventually the top becomes vertical; the time constant of this decrease

(26) is however not equal to that of the homogeneous solutions. When the top is subjected to a horizontal acceleration in addition to gravity, it eventually points opposite to the direction of the net force plus pseudo force.

We end with some suggestions on how to conduct an experiment which will verify our predictions. A considerable number of subtleties have to be taken care of while designing the apparatus, hence we include a brief commentary on these issues. The primary feature to be careful of is that the viscous damping of our model is different from the static friction experienced at a practical rough pivot. An accurate result will be obtained if static friction is minimized and a damping is externally introduced. One way is to use a smooth ball and socket joint as the pivot and immerse the surrounding area in fluid. A still better implementation would be to construct the Kleppnerite top above with the ball made of a light metal like aluminium. A magnetic field can be set up in the region so that Lenz's law serves to retard the pivot's motion. Another fact which must be kept in mind is that the motor driving the top must produce a constant torque output to counter the frictional effect of the pivot. It should not tend towards a constant speed mode of operation, which will result in unwanted $o$ axis torques. For an uncontrolled motor, the series field DC motor in the flux weakening regime is the best choice as its torque speed curve is nearly flat at high speeds. If motor control algorithms are employed, the induction motor with direct torque control will be the ideal candidate.

<p style="text-align:center">*  *  *  *  *</p>

# ACKNOWLEDGEMENT

I am grateful to KVPY, Government of India, for a generous Fellowship.